\documentclass[preprint,12pt]{elsarticle}

\usepackage{amssymb}
\usepackage[utf8]{inputenc}
\usepackage[T1]{fontenc}
\usepackage[english,german]{babel}
\usepackage[stable]{footmisc}
\usepackage[english]{hyperref}
\usepackage{multirow}
\usepackage{booktabs}
\usepackage{graphicx}
\usepackage{caption}
\usepackage{float}
\usepackage{units}
\usepackage{upgreek}
\usepackage{mathtools}
\usepackage{subfigure}
\usepackage{paralist}
\usepackage{xcolor}
\usepackage{tikz}
\usepackage[fixlanguage]{babelbib}
\usepackage{pifont}
\usepackage{cases}
\usepackage{comment}
\usepackage{enumerate}
\usepackage{dcolumn}% Align table columns on decimal point
\usepackage{bm}% bold math
\usepackage{ulem}

\journal{Physica A}

\begin{document}
\selectlanguage{english}

\begin{frontmatter}



\title{Microbiome abundance patterns as attractors and the implications for the inference of microbial interaction networks}

\author[inst1]{Isabella-Hilda Mendler}
\author[inst1]{Barbara Drossel}
\author[inst2]{Marc-Thorsten Hütt}

\affiliation[inst1]{organization={Institute for condensed matter physics, Technical University of Darmstadt},%Department and Organization
            addressline={Hochschulstr. 6}, 
            city={64289 Darmstadt},
            %postcode={64289}, 
            %state={State One},
            country={Germany}}

\affiliation[inst2]{organization={Computational Systems Biology, Department of Life Sciences and Chemistry, Jacobs University Bremen},%Department and Organization
            addressline={Campus Ring 1}, 
            city={28759 Bremen},
            %postcode={28759}, 
            %state={State Two},
            country={Germany}}

\begin{abstract}

Inferring microbial interaction networks from abundance patterns is an important approach to advance our understanding of microbial communities in general and the human microbiome in particular. Here we suggest discriminating two levels of information contained in microbial abundance data: (1) the quantitative abundance values and (2) the pattern of presences and absences of microbial organisms. The latter allows for a binary view on microbiome data and a novel interpretation of microbial data as attractors, or more precisely as fixed points, of a Boolean network. 

Starting from these attractors, our aim is to infer an interaction network between the species present in the microbiome samples. To accomplish this task, we introduce a novel inference method that combines the previously published ESABO (Entropy Shifts of Abundance vectors under Boolean Operations) method with an evolutionary algorithm. The key idea of our approach is that the inferred network should reproduce the original set of (observed) binary
abundance patterns as attractors.

We study the accuracy and runtime properties of this evolutionary method, as well as its behavior under incomplete knowledge of the attractor sets. Based on this theoretical understanding of the method we then show an application to empirical data. 

\end{abstract}




\end{frontmatter}


\section{Introduction \label{sec:intro}}

Microorganisms such as bacteria do not live in isolation, but form complex communities \cite{faust2012microbial}. Species that are part of such a microbial community participate in mutualistic and antagonistic interactions, for instance by cooperating to form a biofilm, or by competing for nutrients. 

Since microbiota and the interactions between their members play a crucial role in the health of their host, 
analyzing microbial abundances has received widespread attention during the last decade \citep{methe2012framework}. For various types of microbial communities, ranging from soil samples to the human skin, mouth or gut, sequencing-based abundance estimation of microbial taxa has become a widespread tool to gather information on the underlying ecosystem. The immense medical relevance and clinical potential of microbiome analysis is becoming more and more apparent \citep{clemente2012impact,brown2013translating,durack2019gut,manor2020health}.

In spite of this relevance, we are still lacking a deep theoretical understanding of microbiome patterns. An early -- and strongly criticized -- attempt has been the hypothesis of distinct microbiome states, the so-called enterotypes \citep{arumugam2011enterotypes}. However, the definition of a microbiome state itself is a challenging problem \citep{garcia2019robust}.  

An important step towards a theoretical understanding of microbiome patterns consists in 
estimating the underlying microbial interaction networks from abundance patterns.  The challenges of extracting reliable microbial interaction networks from abundance patterns have been summarized in \citet{rottjers2018hairballs}, and more recently by \citet{matchado2021network}.

There is a sharp difference in the type of systemic insight provided by (continuous) abundance patterns and (binary) patterns of presences and absences. When one analyzes a career fair at a university, the companies present or absent reveal information about the ties in education and research this university has to companies, while the number of people the company has sent to the career fair (the 'abundance') is rather informative about the size of the company itself or their future hiring ambitions. 

Similarly, in a transcriptomic data set, the absolute expression level of a gene is often indicative of the gene product's function: Typically the expression levels of (genes encoding) transcription factors are much smaller than those of metabolic enzymes, while the "on" and "off" pattern of genes expression is in many cases rather informative about the underlying regulatory networks \citep{liang1998reveal,bornholdt2005less}. \\

The challenge of evaluating both types of information simultaneously has been addressed for example in \citet{prost2021zero}, where the statistical model accounts for extremely sparse abundance data with furthermore broad distributions of the non-zero values. 

Looking at the microbiome from a binary perspective, in addition to the standard 'abundance pattern' approach, has the advantage of differentiating between these distinct types of information. Additionally, it paves the way towards new mathematical approaches of studying the microbiome by viewing microbiome data as attractors of Boolean networks. 

In fact, most inference methods consider only structural network properties \citep[see, e.g.,][]{vidanaarachchi2020imparo,nagpal2020metagenonets}. Only a few recent approaches have taken into account that  the microbiome is a dynamical network and that the measured data should represent steady states of such a network \citep{xiao2017mapping,claussen2017boolean}.

In an earlier investigation \citep{claussen2017boolean} we adapted a presence/absence analysis to the phenomenon of microbiome variability. One of the surprising findings was the large number of systematic and positive interactions, which complement the dominant negative interactions reported in the literature before. This systematic contribution of low-abundance taxa (the 'rare biosphere' \citep{heinken2015systematic}) to the interaction pattern of microbial organisms in the human gut emphasizes their relevance to the metabolic function of the whole system.  

Currently the vast majority of available data sets is still of the type of abundance 'snapshots'. With the availability of more and more longitudinal data (i.e., time series of such abundance patterns), more sophisticated network inference methods can be developed (e.g., dynamic or time-varying network modeling; see \citet{garcia2020can}). 
Methods solely based on correlations and other simple statistical associations have been criticized recently \citep{blanchet2020co}. 

The original ESABO method \citep{claussen2017boolean} also essentially relied on co-occurrence patterns. Here we go one step further and use the ESABO network only as a starting point for the network inference process and then -- via simulated evolution -- require the network also to reproduce the original set of (observed) binary abundance patterns as attractors. We show that this evolution-enhanced ESABO method produces reconstructed networks that have a high resemblance with the original ones. By investigating the method in situations where data do not show all attractors, we find a relationship between the percentage of known attractors and the average accuracy achieved by the reconstruction method.

\newpage
\section{Methods}

\subsection{Entropy Shifts of Abundance vectors under Boolean Operations (ESABO)}

The Entropy Shifts of Abundance vectors under Boolean Operations (ESABO) method is a method for the inference of microbial interaction networks from microbial abundance data.
It was originally introduced in 2017 by Claussen et al. \cite{claussen2017boolean}.
The special feature of this method is that only binary abundance vectors (i.e. the presence or absence of a microbial species in microbiome samples) are considered, which makes it particularly suitable for the investigation of the low-abundance segment of the microbiome.
In general, the ESABO method evaluates the information content of pairs of binary abundance vectors, when combined via Boolean operations. 

The ESABO method starts from a Boolean data matrix $A \in \mathbb{B}^{N_{A} \times N}$ with $\mathbb{B}=\{0;1\}$, where each row vector of the matrix represents a sample and each column vector contains the abundances of a certain species in the different samples. 

The ESABO score for two species $i$ and $j$ is calculated by taking their abundance vectors $\vec{b}_i$ and $\vec{b}_j$ and performing the following four steps \cite{claussen2017boolean}:

\begin{enumerate}[{(1)}]
    \item The logical AND operation is applied component-wise to the two abundance vectors $\vec{b}_i$ and $\vec{b}_j$, i.e.
        \begin{equation*}
            \left(\vec{x}_{ij}^\text{AND}\right)_k = \left(\vec{b}_i\right)_k \text{AND} \left(\vec{b}_j\right)_k.
        \end{equation*}
    \item The entropy of the resulting vector $\vec{x}_{ij}^\text{AND}$ is calculated by 
        \begin{equation*}
        H(\vec{x}_{ij}^\text{AND}) = -\sum_{l \in \{0,1\}} p_l\left(\vec{x}_{ij}^\text{AND} \right) \log\left( p_l\left(\vec{x}_{ij}^\text{AND} \right) \right),
        \end{equation*}
        with 
        $p_l\left(\vec{x}_{ij}^\text{AND} \right)$ being the relative frequency of the entry $l\in \{0,1\}$ in the vector $\vec{x}_{ij}^\text{AND}$.
    \item The entries of the abundance vector $\vec{b}_j$ are randomly permuted, leading to a new vector $\vec{b}_j^*$. 
    Then, the AND operation between the vectors $\vec{b}_i$ and $\vec{b}_j^*$ is performed and the entropy $H(\vec{x}_{ij^*}^\text{AND})$ of $\left(\vec{x}_{ij^*}^\text{AND}\right)_k = \left(\vec{b}_i\right)_k \text{AND} \left(\vec{b}_j^*\right)_k$ is calculated. \\
    This step is repeated $R=1000$ times, in order to obtain a distribution of entropy values (from shuffled versions of $\vec{b}_j$), which serves as a null model.
    \item The obtained entropy mean $\mu$ and standard deviation $\sigma$ are used to calculate the z-score   
        \begin{equation}
           Z_{ij}= \frac{H(\vec{x}_{ij}^\text{AND})-\mu}{\sigma},
    \end{equation}
    which corresponds to the so-called ESABO score for the two species $i$ and $j$. 
\end{enumerate}

For a positive interaction between species $i$ and $j$ we expect the ESABO score to be positive,  and for a negative interaction between $i$ and $j$ we anticipate a negative  ESABO score. 

In order to obtain the reconstructed network, an ESABO-score threshold $\Theta$ needs to be chosen, which determines beyond which absolute z-score value a link is included in the network. In \cite{claussen2017boolean} it was suggested that links with $|Z_{ij}|>1$ should be set. \\

For the present work, we made two changes to the ESABO method. 
First, we sped up the computation of the ESABO scores $Z$ and reduced the randomness of the obtained results by replacing step (3) by an analytical formula for the mean $\mu$ and the standard deviation $\sigma$ of the entropy distribution. This formula is obtained by calculating the entropy for all $N_{A}!$ possible permutations of the entries of $\vec{b}_j$ (see~\ref{sec:appendix_zScore_calc}).  
Second, we improved the ESABO method by interchanging the 0s and 1s in a pair of abundance vectors before performing the logical AND operation if the relative frequency of 1s was higher than 50\% in either of the two considered abundance vectors, i.e., if $p_1\left(\vec{b}_i\right) > 0.5 \, \text{ or } \, p_1\left(\vec{b}_j\right)>0.5$. In this way, we avoided the rare occurrence of large negative z-scores for synergistic interactions, which was an issue in the original version of the ESABO method.

\begin{figure}[h]
\centering
\includegraphics[width=1\linewidth]{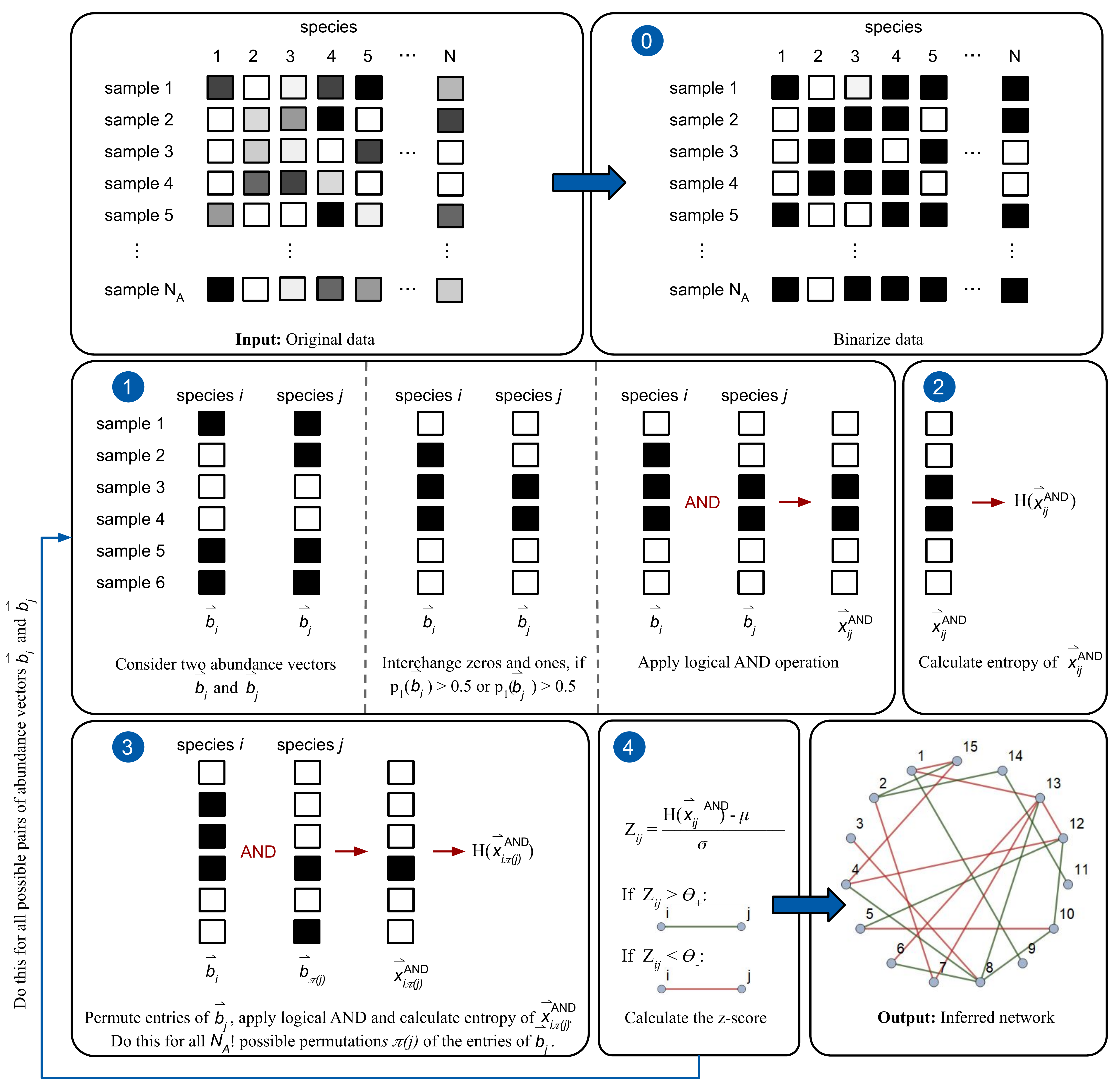}
\caption{
\label{fig:ESABO}
Schematic illustration of the refined ESABO method}
\end{figure}

A schematic illustration of this refined version of the ESABO method is shown in Figure \ref{fig:ESABO}.

\clearpage

\subsection{Generation of simulated data}
\label{subsec:GenSimData}
An advantage of the ESABO method is that it can be easily tested by applying it to artificially generated data.
To this purpose, we first generated a random undirected Boolean network with N nodes, representing $N$ microbial species,
connected by $L_+$ positive and $L_-$ negative links, respectively interactions. 
The nodes in a Boolean network can only assume two different values, namely 1, which means that the considered species is present, or 0, which signifies the absence of the species. 
The change in time of the state of each node is determined by its Boolean update function. We used threshold functions, where the value $s_i$ of each node or species $i$ depends only on the sum of its input signals and is updated at each time step according to
\begin{equation}
s_i(t+1) =
    \begin{cases}
    1, & \sum_{j=1}^N G_{ij} s_j(t) > 0 \\
    s_i(t), & \sum_{j=1}^N G_{ij} s_j(t) = 0 \\
    0, &  \sum_{j=1}^N G_{ij} s_j(t) < 0\\
    \end{cases}.
\label{eq:update_function}
\end{equation}
$G$ is the generalized adjacency matrix of the interaction graph with
\begin{displaymath}
G_{ij} =
\begin{cases}
+1, \quad  \text{for a positive interaction between $i$ and $j$}  \\ 
-1, \quad  \text{for a negative interaction between $i$ and $j$}  \\ 
\phantom{-} 0, \quad \text{if there is no interaction between $i$ and $j$}. 
\end{cases}
\end{displaymath} 

Since we considered undirected networks, $G$ is symmetric ($G_{ij}=G_{ji}$) and we assumed that $G_{ii}=0 \, \forall i$, i.e., we did not consider self-inputs.
Furthermore, all the nodes of the network were updated synchronously. \\

After the generation of the network, we determined its attractors,  using the algorithm described in \cite{Hopfensitz2013}. To ensure that all attractors of the network are found, we updated the network from each of its $2^{N}$ possible states. In the rare case where the attractor of the network is not a fixed point but a cyclic attractor, we chose the first encountered state from the cycle as the recorded steady-state. \\

The attractors are interpreted within the framework of the ESABO method as steady-state microbiome compositions. All or part of these attractors are chosen to represent the 'samples', from which the network is reconstructed by the ESABO method.

Networks that were reconstructed with the ESABO method usually display only part of the attractors that were used to perform the network inference (cf. Fig. \ref{fig:EvolutionResult_AllAttr} left). 

For this reason, we  subjected the reconstructed networks to a simple evolutionary algorithm, which is based on mutation and selection, in order to improve their ability to reproduce the attractors of the original network. The resulting ESABO enhanced evolutionary algorithm is described in the following section. 

\subsection{ESABO enhanced evolution of reconstructed networks}

\begin{figure}[h]
\centering
\includegraphics[width=0.65\linewidth]{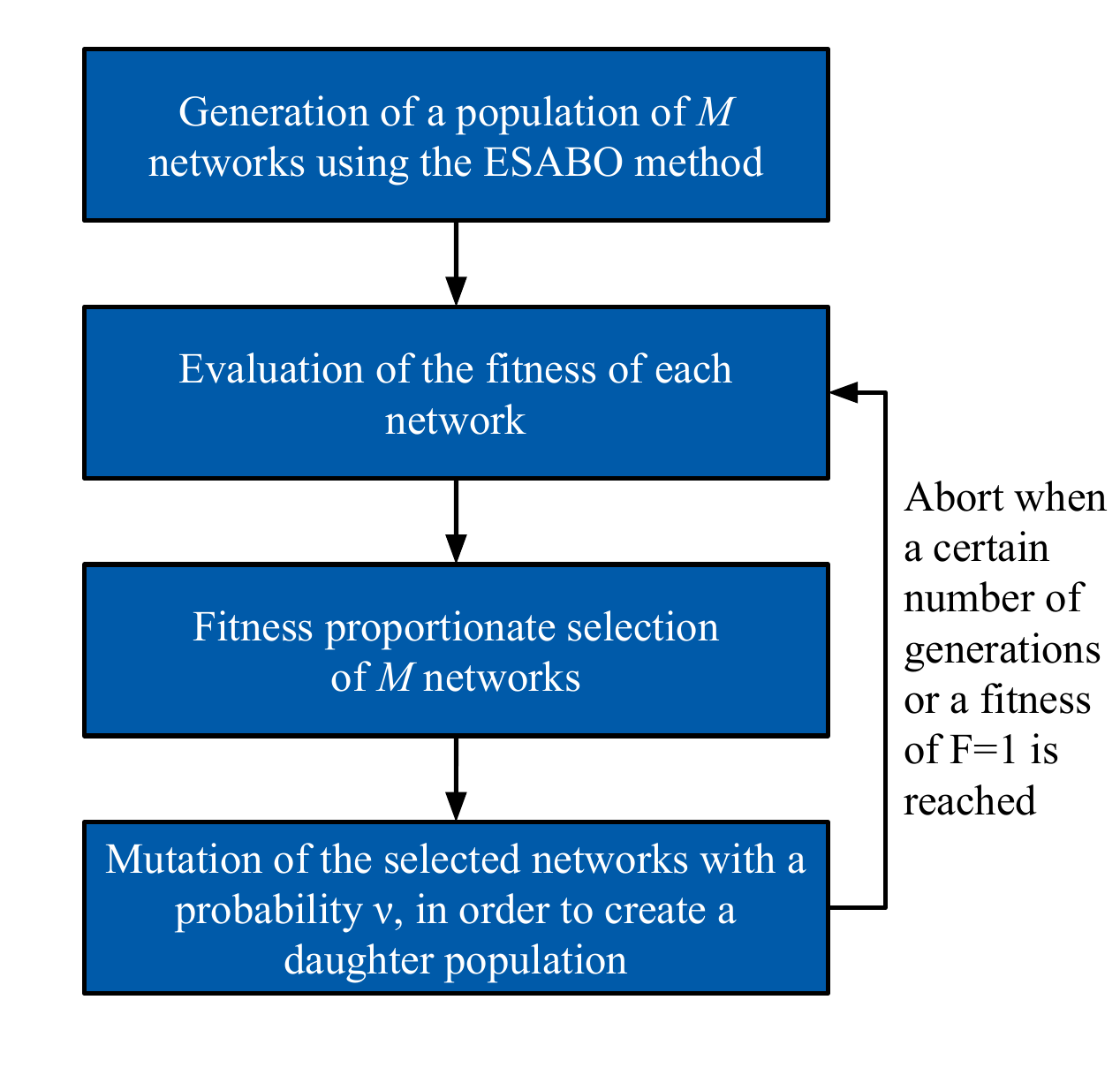}
\caption{Schematic illustration of the ESABO enhanced evolutionary algorithm.
\label{fig:EvoAlgorithm}
}
\end{figure}

The evolutionary algorithm is shown schematically in Figure \ref{fig:EvoAlgorithm}. It consists of the following four key elements: The generation of a population of $M$ networks using the ESABO method, the determination of the fitness of all networks in the population, the fitness proportionate selection of $M$ networks, and the mutation of the selected networks, in order to create the next generation. The implementation of these four elements is described in the following. \\

\newpage
\textbf{Generation of a population of $M$ networks using the ESABO method:}
In order to generate the initial population of $M$ networks, we first calculate the ESABO scores for all $\frac{N \cdot (N-1)}{2}$ possible links in the network.
Then, we construct $M$ networks with an increasing number of links by starting from the network that contains the $L_{min}=10$ links that have the highest absolute z-score values. We add successively  further links according to their z score, until networks with $L_{min}+M-1$ links are obtained. \\

\textbf{Evaluation of the fitness of each network:}
If all attractors of the original network are used for its inference, the fitness of a network corresponds to the Jaccard index
\begin{equation}\label{fitness}
F= \frac{\left|A_\text{ori} \cap A \right|}{\left|A_\text{ori} \cup A \right|}
\end{equation}
 between its attractors $A$ and the attractors $A_\text{ori}$ of the original network, respectively the original samples. The advantage of this performance measure is that it can also be evaluated for biological data, where the real interaction network is unknown.

In the case of real microbiomes, we cannot assume that all the attractors of the underlying interaction network are present in the available samples. Especially attractors with small basin sizes might not be represented in the data.
Therefore, we investigate what happens if some attractors, particularly those with small basin sizes, are not considered during the network inference. 
We generated the desired sample size $N_A$ (taken to be a fixed proportion of the total attractor number) by picking from the original network the $N_A$ attractors with the largest basin sizes.

To assess the fitness of a reconstructed network, we initialized it from up to $ N_\text{ini}=1000$ random initial states 
and iterated the dynamics until an attractor was reached. Whenever we obtained as many attractors as the number of samples, we stopped the search. 
Since the Jaccard index \eqref{fitness} of such a reconstructed attractor set varies with each evaluation of the attractors, we averaged it over $r=100$ runs (unless stated otherwise). This mean Jaccard index was used as the fitness of the network. \\

\newpage
\textbf{Fitness proportionate selection of $M$ networks:}
In order to create a daughter population, we selected networks with a probability proportional to their relative fitness, i.e.,
the weight $W_i$ with which individual i was chosen to be the parent of a given individual of the next generation is
\begin{equation*}
  W_i = \frac{F_i}{\sum_{j=1}^M F_j}. 
\end{equation*} \\

\textbf{Mutation of the selected networks}:
The daughter network was created from the selected parent by making a copy of the parent and applying to it a mutation with the probability $\nu$. 
The following three types of mutations, all of which occur with the same probability, were performed: 

\begin{enumerate}[{(1)}]
    \item Deletion of a randomly selected link.
    \item Addition of a new link to the network (according to its ESABO score): \\
          First, we decided with equal probability whether a positive or a negative link should be added. The sign of a link was determined by the sign of its ESABO score $Z$.
          Then, within the chosen set, we randomly drew a link linearly  weighted by its ESABO score.  
    \item Change of a link: This is a combination of mutation (1) and (2).
\end{enumerate}

Using such an evolutionary algorithm has the great advantage that it is not necessary to (manually) select a certain threshold that defines how many links should be included in the network. This is a crucial task in many other network inference methods like e.g. SparCC \cite{Friedman_SparCC} or SPIEC-EASI \cite{Kurtz_SPIEC-EASI} and can here be avoided.

\newpage
\section{Results}

\subsection{Analysis of simulated data under the assumption that all attractors of a network are known}

First, we tested our method using simulated data, as described in section \ref{subsec:GenSimData}. For this purpose, we considered 40 random networks with N = 15 nodes and $L_{+}=L_{-}=10$ positive and negative links. All these networks had more than 200 distinct attractors and were connected. 

We reconstructed these networks using the ESABO enhanced evolutionary algorithm with 
a population size of $M=50$ and a mutation probability of $\nu = 0.25$.
In the initial population the network with the fewest links had $L_\text{min}=10$ links and the network with the most links had $L_\text{min}+M-1=59$ of them.\\

We evaluated the performance of our method according to two different measures.

First, we evaluated the fitness \eqref{fitness} of the evolved networks. The fitness of a network measures to what extent its attractors match those of the original network. 

Second, we checked how well the network topologies, i.e., the links of the inferred network and those of the original network, agreed. 
This was quantified via the Jaccard index 
\begin{equation*}
    J=\frac{\left|L_\text{ori} \cap L \right|}{\left|L_\text{ori} \cup L \right|}
\end{equation*}
between the links $L$ of the fittest evolved network and the links $L_\text{ori}$ of the original network. \\

Moreover, we compared the ESABO enhanced evolution to two different types of a random evolution, where a completely random link is set during mutation (2), regardless of its ESABO score or expected sign. While the first type of random evolution is starting from networks that were reconstructed with the ESABO method (like in the ESABO enhanced evolution), the second type of random evolution is starting from a population of random networks. \\

\begin{figure}[t]
\centering
\includegraphics[width=0.6\linewidth]{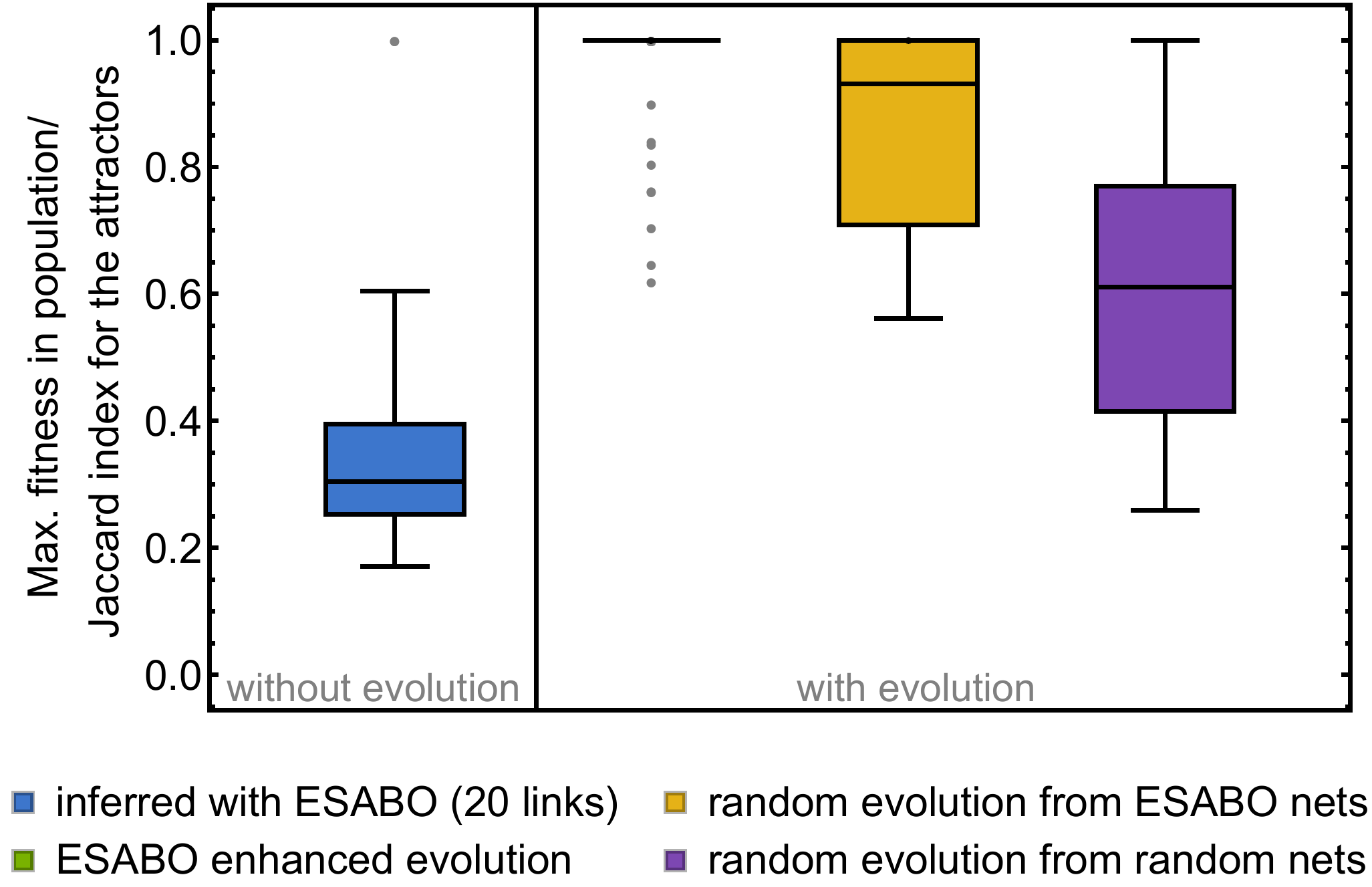}
\caption{
Box plots showing the Jaccard index between the attractors of the original network and the attractors of the inferred or fittest evolved network that was found at some point during evolution. $40$ networks with $N=15$ nodes and $L_{+}=L_{-}=10$ positive and negative links were investigated. The evolution was performed for 10000 generations with $M=50$, $\nu=0.25$ and $L_\text{min}=10$. For the network inference with the ESABO method the 20 links with the highest absolute ESABO scores were set.
\label{fig:EvolutionResult_AllAttr}
}
\end{figure}

\begin{figure}
\centering
\includegraphics[width=\linewidth]{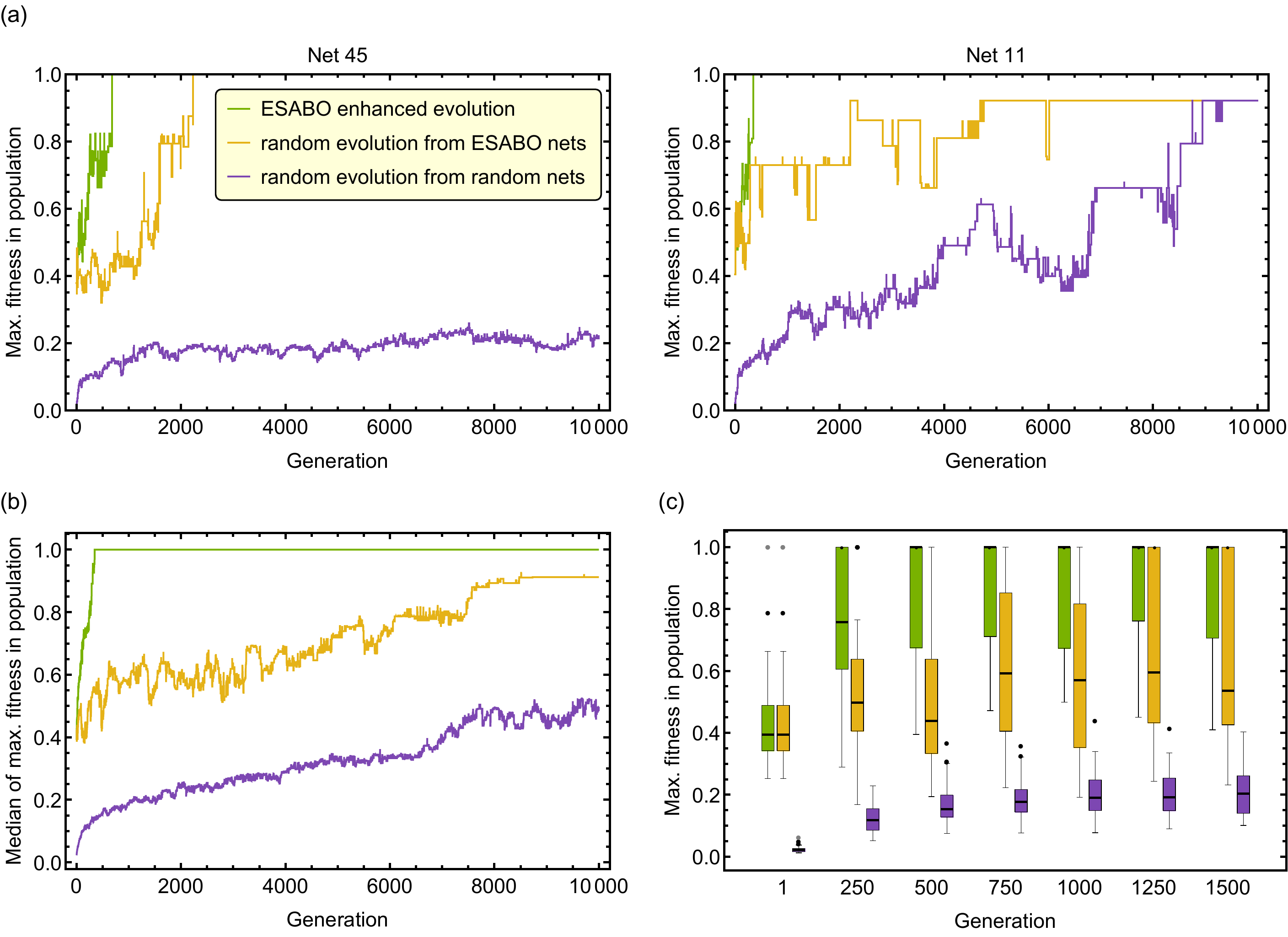}
\caption{
\label{fig:MaxFitness_Evolution}
Evolution of the maximum fitness in the  population for the three different versions of the evolutionary algorithm. \\
(a) Fitness increase in course of the evolution for two exemplary networks. \\
(b) Median of the maximum fitness for the 40 investigated networks in the course of evolution. \\
(c) Fitness distribution for the 40 investigated networks in form of box plots.}
\end{figure} 

\newpage
In general, we obtained with the ESABO enhanced evolutionary algorithm networks that display the same attractors as the original networks, i.e. that have a fitness of $F=1$. This can be seen in
Figure \ref{fig:EvolutionResult_AllAttr}, which shows the maximum fitness value that was obtained for each of the 40 investigated networks at some point during an evolution of 10000 generations. While networks that were solely reconstructed with the ESABO method only have a median fitness of $F=0.3$, networks that were subjected to an ESABO enhanced evolution have a median fitness of $F=1$. The other two evolutionary algorithms perform better than the ESABO method without evolution, but show a broader distribution and a significantly lower median fitness value than the ESABO enhanced evolution. 

The main reason for this is that the ESABO enhanced evolution is much faster than a random evolution. This is shown in Figure \ref{fig:MaxFitness_Evolution}, where the evolution of the maximum fitness in the  population is displayed for the three different versions of the evolutionary algorithm.
While Fig. \ref{fig:MaxFitness_Evolution} (a) shows the evolution of the maximum fitness for two exemplary networks,
Fig. \ref{fig:MaxFitness_Evolution} (b) displays the median of the maximum fitness for the 40 investigated networks and Fig. \ref{fig:MaxFitness_Evolution} (c) shows the fitness distributions in form of box plots.  \\
\newpage
Networks that were evolved using the ESABO enhanced evolution usually show a steep fitness increase (Fig. \ref{fig:MaxFitness_Evolution} (a)) and reach a median fitness of $F=1$ in less than 500 generations (Fig. \ref{fig:MaxFitness_Evolution} (b), (c)), while the fitness increase is considerably slower for both types of the random evolution.
Randomly evolved networks, where the evolution starts from a population of random networks, have even after 10000 generations a significantly lower median fitness of $F \approx 0.5$. \\

\begin{figure}[h]
\centering
\includegraphics[width=0.6\linewidth]{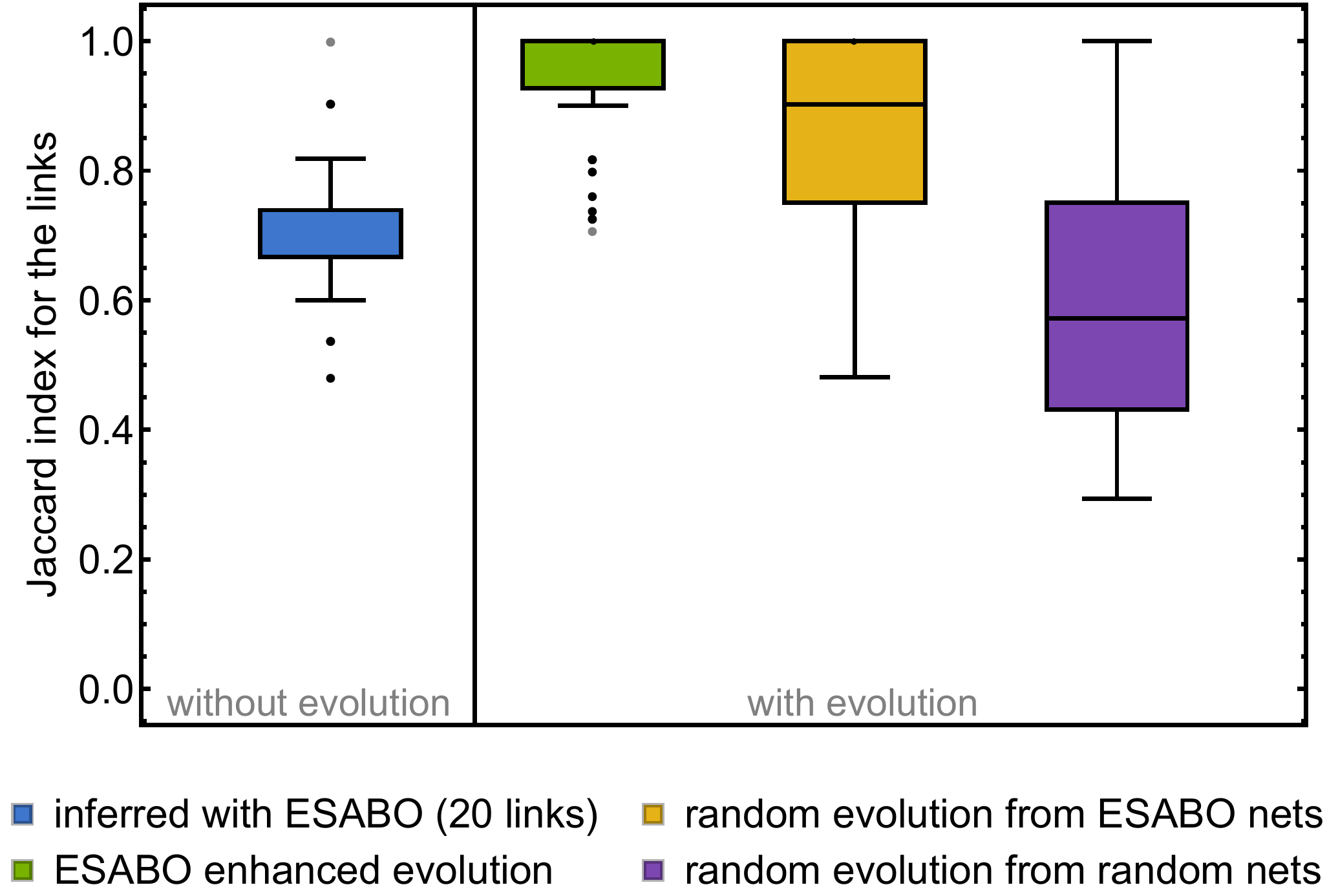}
\caption{
Box plots showing the Jaccard index between the links of the original network and the links of the inferred or fittest evolved network that was found at some point during evolution  (same networks as in Fig. \ref{fig:EvolutionResult_AllAttr}). $40$ networks with $N=15$ nodes and $L_{+}=L_{-}=10$ positive and negative links were investigated. The evolution was performed for 10000 generations with $M=50$, $\nu=0.25$ and $L_\text{min}=10$. For the network inference with the ESABO method the 20 links with the highest absolute ESABO scores were set.
\label{fig:ReconstructionResult_AllAttr}
}
\end{figure}

Furthermore, the networks that were evolved using the ESABO enhanced evolution do not only have a similar dynamics to the original networks, but they are also topologically very similar to them. 
This can be seen in Figure \ref{fig:ReconstructionResult_AllAttr}, where the Jaccard index between the links of the original network and the links of the fittest inferred network that was found at some point during evolution is shown for the $40$ investigated networks.
If we compare the reconstruction quality, we find that networks that were inferred with the ESABO method have a median Jaccard index of $J=0.74$, whereas networks that were subjected to the ESABO enhanced evolution have a significantly higher median Jaccard index of $J=1$.

\subsection{Analysis of simulated data under the assumption that only a part of the attractors is known}

\begin{figure}[h]
\centering
\includegraphics[width=0.95\linewidth]{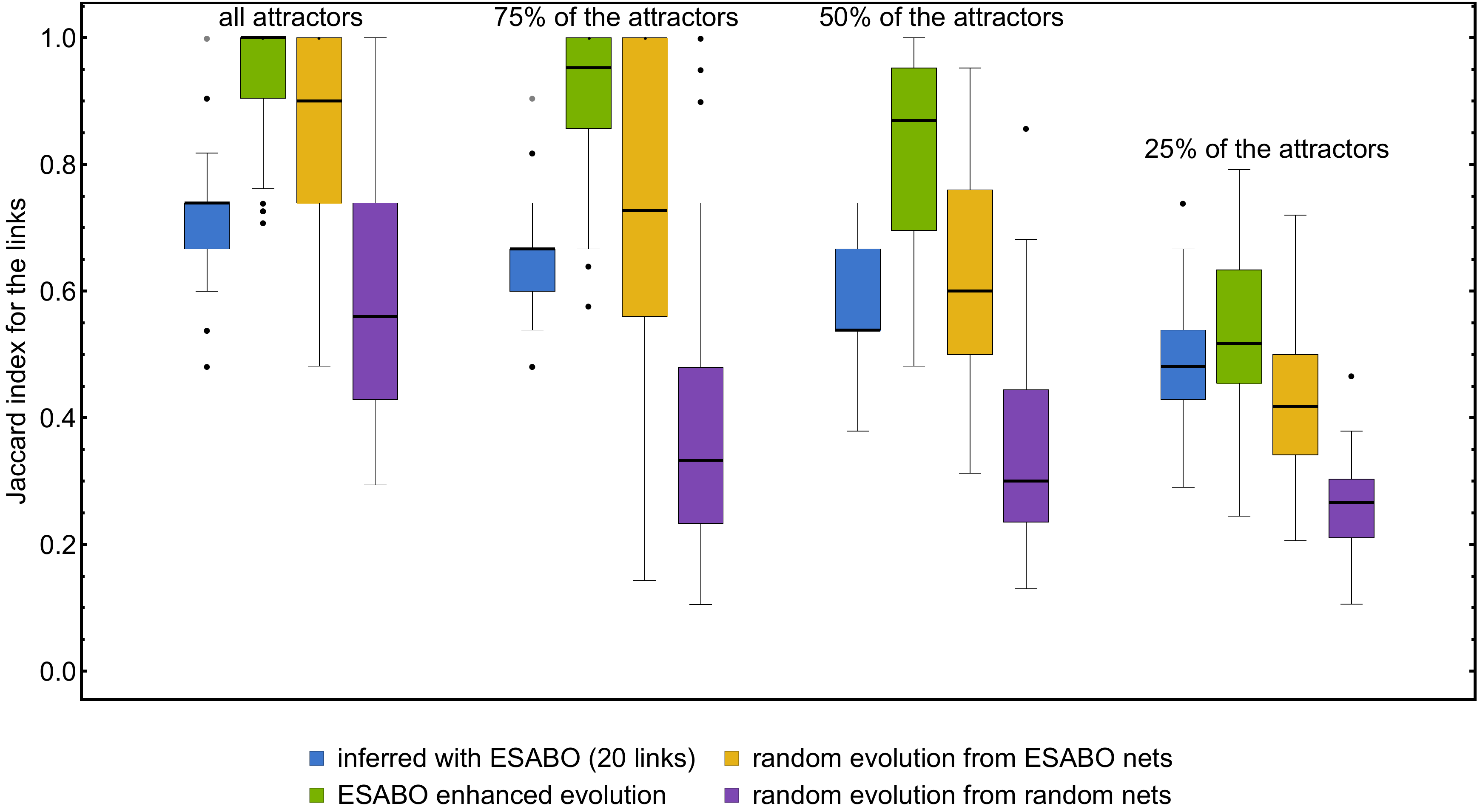}
\caption{
Box plots showing the Jaccard index between the links of the original network and the links of the inferred or fittest evolved network. $40$ networks with $N=15$ nodes and $L_{+}=L_{-}=10$ positive and negative links were investigated. The evolution was performed for 10000 generations with $M=50$, $\nu=0.25$ and $L_\text{min}=10$.
\label{fig:ReconstructionResult_NotAllAttr}
}
\end{figure}

We consider again the 40 random networks from the previous section and only use a certain percentage of their attractors (always those with the largest basin sizes) to reconstruct the networks with the ESABO enhanced evolutionary algorithm. 

Figure \ref{fig:ReconstructionResult_NotAllAttr} shows the reconstruction quality, in terms of the Jaccard index between the links of the original network and those of the reconstructed network, if either all attractors, 75\%, 50\% or 25\% of them were used for the network inference. 
As we can see, the ESABO enhanced evolution is always superior to the simple ESABO method as well as to the other evolution types. It still works very well if only 50\% of the attractors are used as an input for the network inference. In this case, the ESABO enhanced evolution achieves a median reconstruction quality of $J=0.87$. 
In the case where only 25\% of the attractors are considered for the reconstruction of the network, the ESABO enhanced evolution shows a large drop in the inference performance and only reaches a median inference quality of $J=0.52$.\\

Before applying our inference method to real biological abundance data, we expanded our investigation to larger random networks with $N=22$ nodes, 
in order to test our algorithm for the same number of nodes, respectively classes, as present in the investigated abundance data (see section \ref{subsec:saliva_microbiome}).
Since these networks have a much larger state space ($2^N=2^{22}=4194304$) than networks with 15 nodes ($2^{15}=32768$), we increased the maximum number of initial states that were used to find the attractors of a network during the evolutionary process to $N_\text{ini}=100000$. To keep computation times within reasonable limits, we reduced the repetitions of the Jaccard-index calculation to $r=10$, and we modified the evolutionary algorithm such that we always kept the fittest network, i.e., we copied it to the next generation without a mutation. Fitness-proportionate selection with a (possible) subsequent mutation was only applied to the other $M-1$ networks in the population. 

\begin{figure}
\centering
 \includegraphics[width=0.9\linewidth]{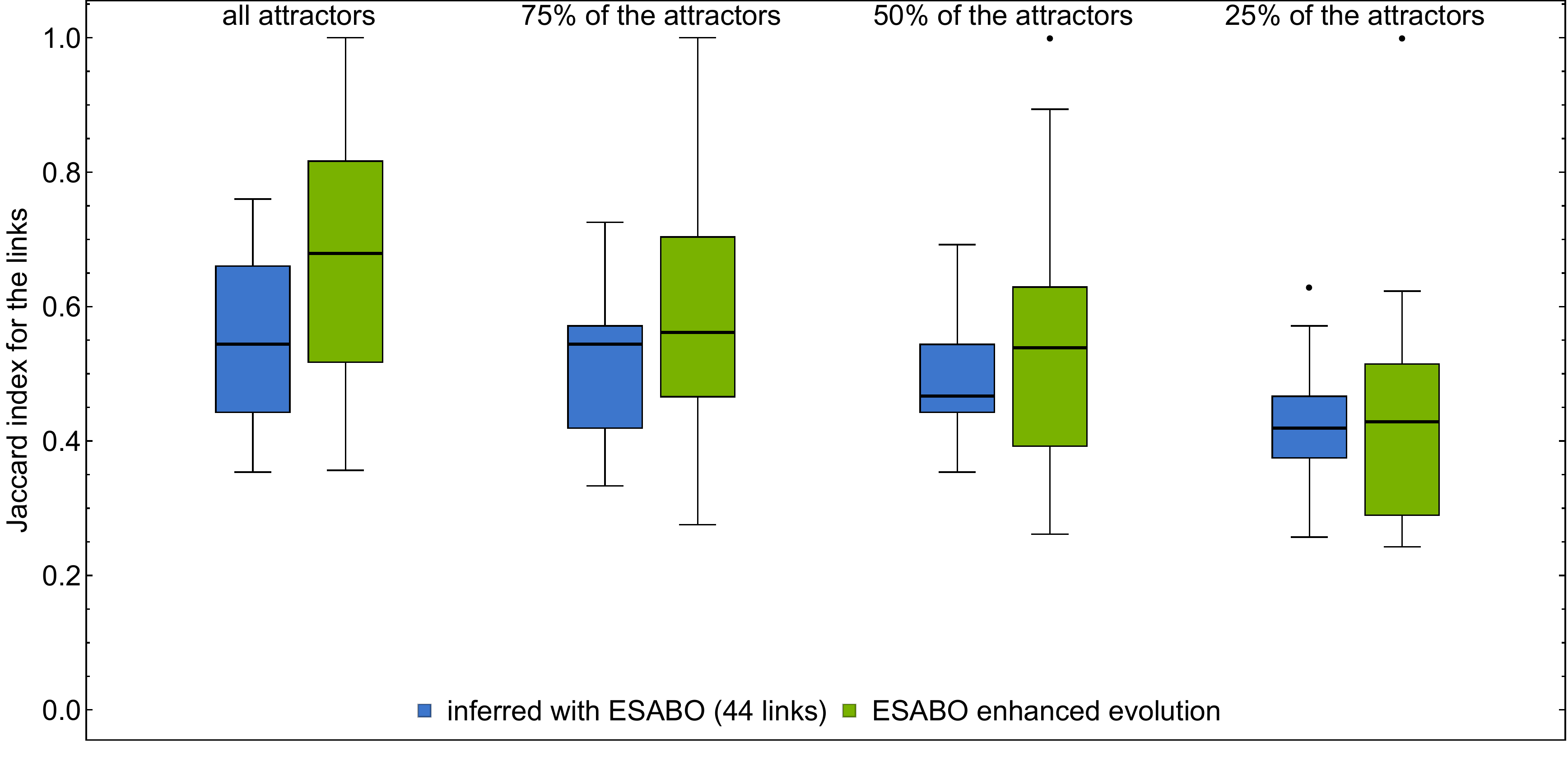}
\caption{Box plots showing the Jaccard index between the links of the original network and the links of the inferred or fittest evolved network. $25$ random networks with $N=22$ nodes and $L_{+}=L_{-}=22$ positive and negative links were investigated. The evolution was performed for 2000 generations with $M=50$, $L_\text{min}=10$, $\nu=0.25$, $N_\text{ini}=100000$ and $r=10$.
\label{fig:Jaccard_links_22nodes}
}
\end{figure}

As shown in Fig.~\ref{fig:Jaccard_links_22nodes},  the ESABO enhanced evolution yields better results than the simple ESABO method also for these larger networks. Even after a short evolution of only 2000 generations it results in higher mean Jaccard-index values, i.e., in networks that are topologically more similar to the original networks than the networks that were inferred with the simple ESABO method. 
A longer evolution time would most likely improve the results, but is associated with long computation times, %due to the attractor search 
especially for the case where all attractors are taken into account. \\

\begin{figure}
\centering
\includegraphics[width=0.9\linewidth]{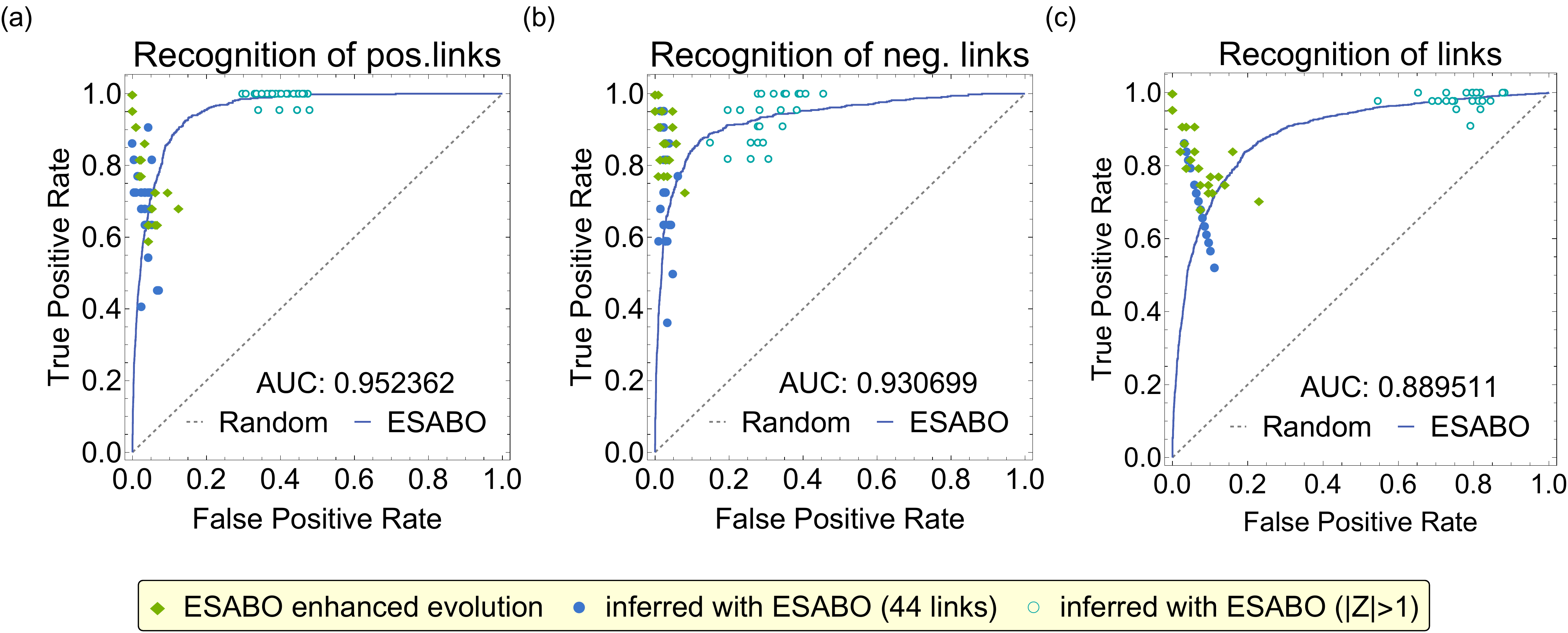}
\caption{Comparison of inference quality between the ESABO enhanced evolution and the original ESABO method using receiver operating characteristics (ROC) curves. 
In order to create the ROC curves for the ESABO method, we merged the link predictions of all the 25 investigated networks from Fig. \ref{fig:Jaccard_links_22nodes} into one large set and ranked them according to their ESABO score. Furthermore, we evaluated the true positive and the false positive rate for networks that were either inferred with the ESABO enhanced evolution or the simple ESABO method. For networks that were inferred with the simple ESABO method, we either chose to set the 44 links with the highest absolute ESABO-score values or all the links with an ESABO score $|Z|>1$. For more details, see \ref{sec:appendix_ROC}. 
The AUC value indicates the area under the ROC curve.
\label{fig:ROC}
}
\end{figure}

The superiority of the ESABO enhanced evolution over the simple ESABO method is also confirmed by examining the true positive rate (TPR) and the false positive rate (FPR) of the evolved or inferred networks as well as the receiver operating characteristics (ROC) graph of the simple ESABO method (see Fig. \ref{fig:ROC} and  \ref{sec:appendix_ROC}). As we can see, the simple ESABO method already provides a good inference quality with area under the ROC curve (AUC) values of $0.95$, respectively $0.93$ for the recognition of positive, respectively negative links. However, it is less successful in distinguishing whether a link is present or not, regardless of its sign (AUC value of $0.89$ for the recognition of links in general).
This is also reflected by the fact, that networks which were inferred with the simple ESABO method setting all the links with an ESABO score $|Z|>1$ as suggested in \cite{claussen2017boolean}, usually have a TPR close to $1$, but a relatively high FPR ($>0.5$ for the recognition of links).  If we only set the 44 links with the highest absolute ESABO-score values, the FPR decreases considerably, but the TPR decreases as well and, most importantly, we used our prior knowledge of the number of links present in the original network. The ESABO enhanced evolution generally yields higher true positive rates at comparable false positive rates as the simple ESABO method, where the 44 links with the highest absolute ESABO-score values were set. Moreover, it achieves this result without any prior knowledge on the number of edges present in the original network. \\

\begin{figure}
\centering
\includegraphics[width=0.9\linewidth]{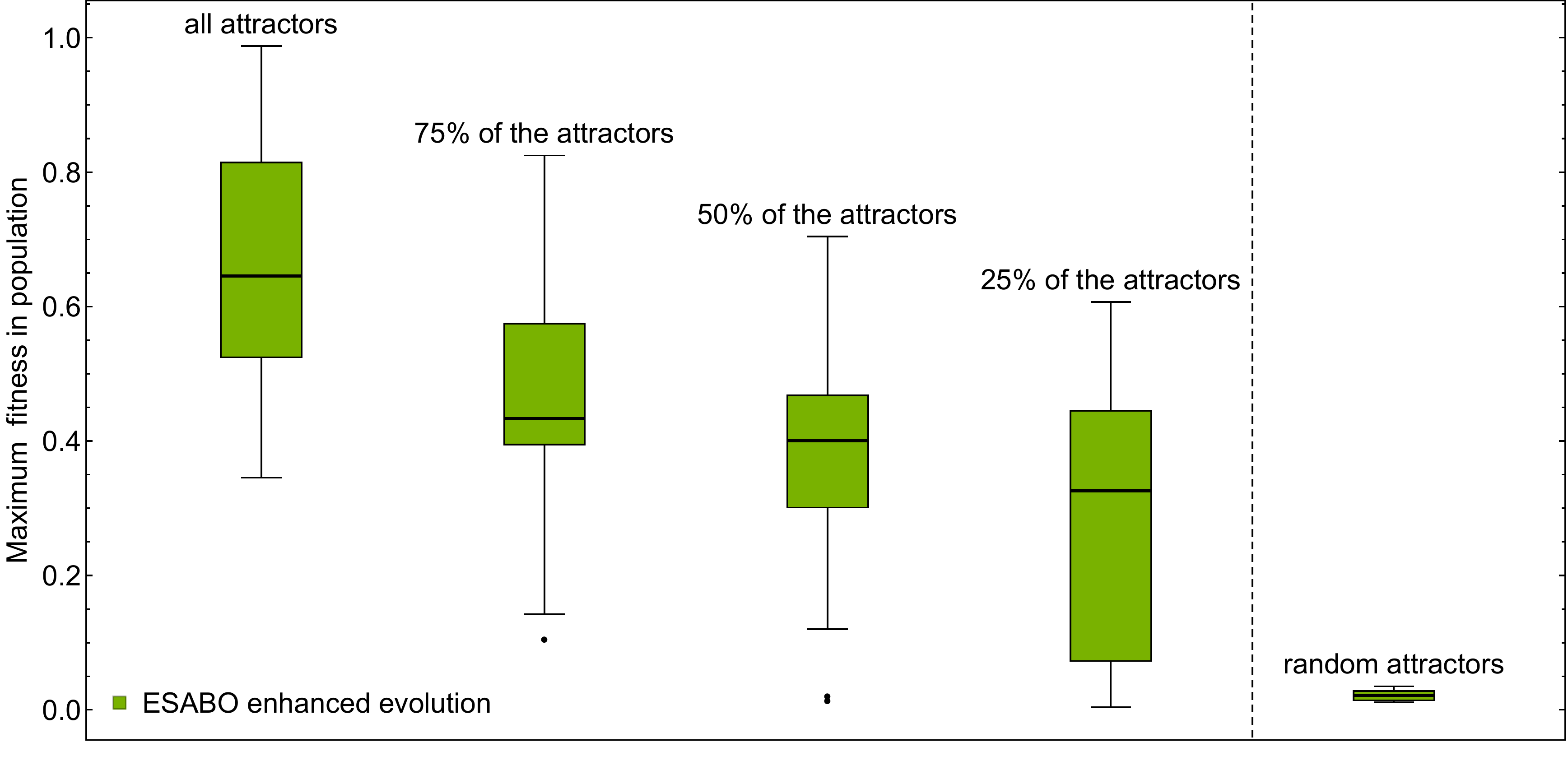}
\caption{
Box plots showing the fitness of the evolved networks from Fig. \ref{fig:Jaccard_links_22nodes} compared to the fitness of 25 networks which were inferred from 138 different random attractors. 
%(that do not belong to an actual network). 
To generate a random attractor, we randomly chose for each of its 22 entries the value 1 or 0 with equal probability. 
\label{fig:Fitness_22nodes}
}
\end{figure}

Figure \ref{fig:Fitness_22nodes} shows the fitness of the evolved networks in comparison to the fitness of randomly assembled attractors that do not belong to an actual Boolean network. 
Although, as expected, the maximum fitness achieved during evolution decreases with decreasing percentage of attractors used for the reconstruction, even in the case where only 25\% of the original attractors were used as an input to the inference method, the reached fitness values ($F_\text{median} \approx 0.33$) are at least an order of magnitude larger than those of randomly assembled attractors ($F_\text{median} \approx 0.02$).

\clearpage
\subsection{Analysis of the human salivary microbiome composition}
\label{subsec:saliva_microbiome}

Finally, we applied the ESABO method to biological abundance data that was derived from 16S ribosomal RNA gene sequences as part of the Human Microbiome Project (HMP). This data was processed by the software package QIIME (Quantitative Insights Into Microbial Ecology), and the resulting operational taxonomic unit (OTU) or phylotype counts were made available at \url{https://www.hmpdacc.org/hmp/HMQCP/}. 

In the following, we only consider the data obtained from the 16S variable region 3-5 (V35). We chose to base our analysis on saliva samples 
since the salivary microbiome of an adult human is rather stable over time \cite{belstrom2016, Lazarevic2010} and therefore can be considered to be an attractor state.

Co-occurrences were analyzed on the class level, and a binarization threshold of 1 was used. Classes that occurred in each of the samples were not considered, since our method requires variation in the presence of a species to predict its interactions.
The results of our investigation are shown in Figure \ref{fig:Saliva_results}. 

As we can see, the ESABO enhanced evolution leads to a relatively large increase in fitness (Fig.~\ref{fig:Saliva_results} (a)). While the fittest network that was inferred with the simple ESABO method (generation 0 of the evolution) has only a fitness of $F\approx0.01$, the evolved network has a fitness of $F\approx0.27$ (after an evolution period of 3000 generations). 
Furthermore, the fitness of the evolved network, which was inferred from real biological data (138 different Boolean samples), is significantly higher than the fitness of networks that were inferred from randomly chosen attractors (not belonging to an actual Boolean network). These networks only reach a median fitness of $F\approx0.02$ after an evolution of 3000 generations. 
This means that our evolutionary inference method recognizes that the biological samples are not random, but do in fact belong to an underlying network.

The largest connected component of the presumed underlying saliva network that was found after an evolution of 3000 generations is shown in Fig. \ref{fig:Saliva_results} (b).
The complete inferred network (with 5 completely unconnected nodes) can be seen in the supplement. We observe that the inferred network, respectively its largest connected component, has considerably more negative links (34 negative links)  than positive ones (9 positive links).
Although it seems interesting to compare this network to other inferred interaction networks for the salivary microbiome, such a comparison is not very useful, since our analysis focuses on the lowly abundant species (we did not consider the classes \textit{Bacilli}, \textit{Bacteroidia}, \textit{Betaproteobacteria}, \textit{Clostridia} and \textit{Gammaproteobacteria}, which occurred in each of the samples), while many other studies (like e.g. \cite{faust2012microbial}) mainly predict interactions between highly abundant phyla or classes (see also the discussion in \cite{claussen2017boolean}).

Instead, we take a closer look at the dynamic properties, respectively the attractors, of the resulting network.
If we compare the original attractors, i.e., the binarized samples, to the attractors of the reconstructed network, we find that they are very similar. Both sets have 13 attractors in common and the remaining 30 samples have a very small Hamming distance of $h \leq 3$ to the attractors of the reconstructed network (see Fig. \ref{fig:Saliva_results}(c)-(d)). 
Furthermore, a comparison of Figure \ref{fig:Saliva_results}(a) with Figure \ref{fig:Fitness_22nodes} suggests that the currently available data cover less than 50 percent of the attractors of the system.

\begin{figure}[h]
\centering
\includegraphics[width=\linewidth]{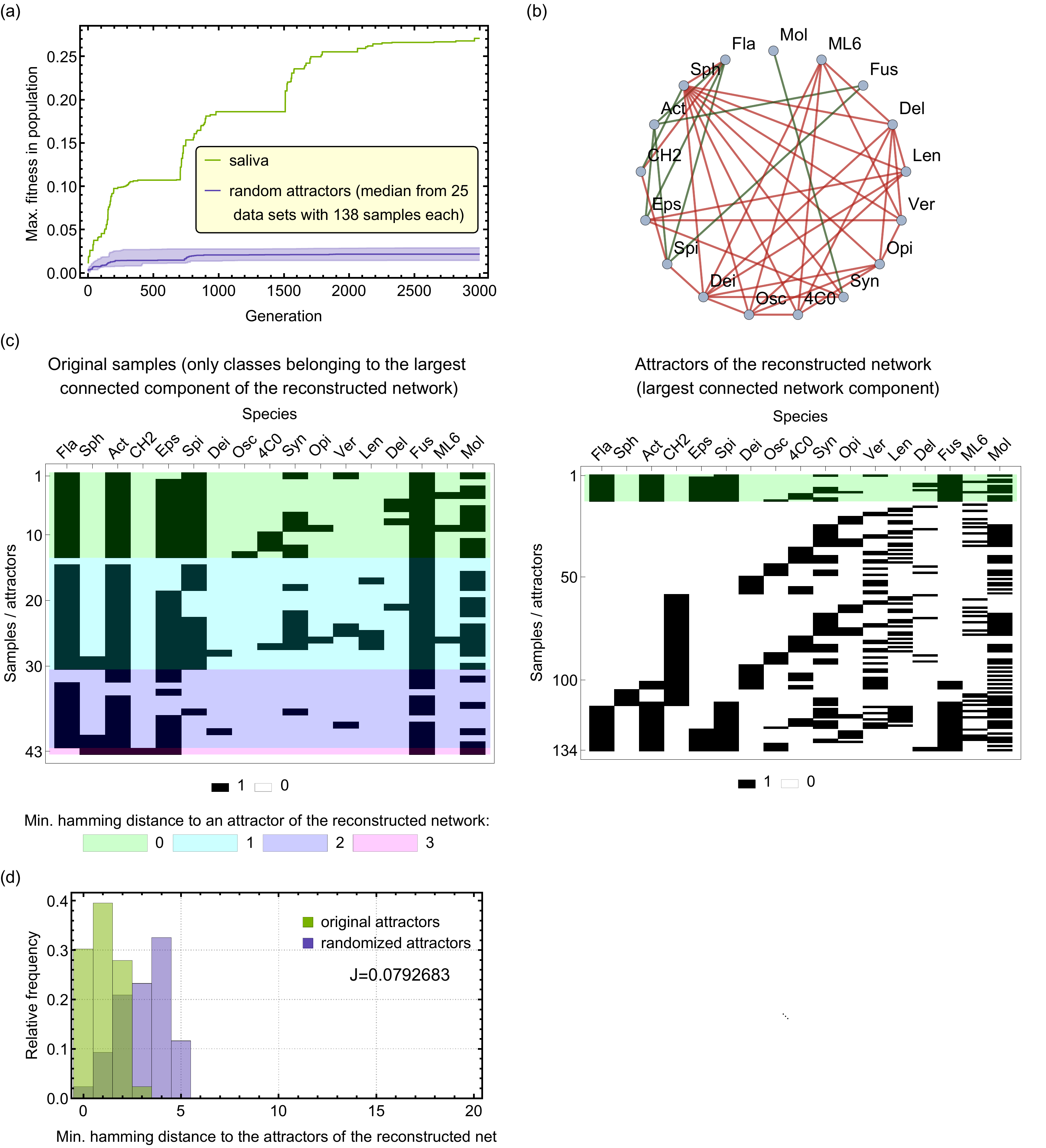}
\end{figure} 
\clearpage

\begin{figure}[H]
\captionof{figure}{
Analysis of the human salivary microbiome \\
(a) Evolution of the maximum fitness for the reconstructed biological network in comparison to the evolution of the median fitness of 25 networks which were inferred from 138 random attractors that do not belong to an actual Boolean network (cf. Fig. \ref{fig:Fitness_22nodes}). The shaded area indicates the area between the 10\% and 90\% quantile.\\
(b) Largest connected component of the  reconstructed saliva network (fittest network that was found after an evolution of 3000 generations). The other 5 nodes that are not part of this component, were not connected at all, i.e. they do not have any links to other nodes. Abbreviations:
Fla: \textit{Flavobacteria}, Sph: \textit{Sphingobacteria}, Act: \textit{Actinobacteria}, CH2: \textit{CH21}, Eps: \textit{Epsilonproteobacteria}, Spi: \textit{Spirochaetes}, Dei: \textit{Deinococci},  Osc: \textit{Oscillatoriophycideae}, 4C0: \textit{4C0d-2}, Syn: \textit{Synergistia}, Opi: \textit{Opitutae}, Ver: \textit{Verrucomicrobiae}, Len: \textit{Lentisphaerae}, Del: \textit{Deltaproteobacteria}, Fus: \textit{Fusobacteria}, ML6: \textit{ML615J-28}, Mol: \textit{Mollicutes}. \\
(c) Comparison of the original attractors (derived from observed abundance patterns) to the attractors of the reconstructed network. In both cases, we only considered the classes that were present in the largest connected component of the reconstructed network. For each sample, the minimum Hamming distance to an attractor of the reconstructed network is indicated by its color. Samples colored in green are reproduced as attractors of the reconstructed network. \\
(d) Histogram showing the minimum Hamming distances of the original attractors (samples) or randomized versions of these attractors to the attractors of the reconstructed network. The  randomization of an attractors was performed by a permutation of its entries. J indicates the Jaccard index between the attractors of the reconstructed network and the original samples from (c).
\label{fig:Saliva_results}}
\end{figure}
\clearpage

\section{Discussion}

Based on a previously introduced network inference method \citep{claussen2017boolean}, which evaluates the co-occurrences of zeros and ones at pairs of nodes, we formulate here a novel approach by evolving the inferred network such that the overlap of attractor sets with respect to the original attractor set (the microbiome data) is maximized. In this way we achieve two goals that are of fundamental importance to a formal description of the microbiome: (1) We infer networks that are by design capable of reproducing the experimental data on a binarized level. (2) We study in detail how this evolutionary inference method is affected by incomplete information on all possible dynamical states (i.e., when only a certain percentage of available attractors have been experimentally observed). 

The original ESABO method feeds in two ways into the evolutionary approach implemented here. The ESABO network serves as a starting point for the evolution, and in addition, the ESABO scores permit prioritizing the edges in the network during the evolution. This dramatically accelerates the simulated evolution. 

The focus on the information contained in the presence/absence patterns of microbial species is not intended as an alternative, but rather as a complement, to abundance-based inference methods: As outlined in the introduction, these two levels of information contained in a microbiome dataset reveal markedly different systemic properties. Based on the numerical experiments performed in our investigation and the discussion of the ESABO method in \citet{claussen2017boolean}, we believe that the Boolean perspective emphasizes rare microorganisms and their contribution to the microbial community, as well as the \textit{intrinsic} interactions among microorganisms, while the abundance perspective puts a stronger emphasis on the dominant microorganisms (with \textit{Firmicutes} and \textit{Bacteroidetes} being prominent examples; see \citet{mariat2009firmicutes}) and is a more reliable indicator of \textit{external} stimuli affecting large parts of the community. 

Focusing on the binary level of information in microbiome compositions (i.e., the presence/absence view on microbial abundance patterns) allows us --  under the assumption that these binary states represent stable attractors -- to relate the microbial interaction network with microbiome states (attractors) in an essentially parameter-free way. 

\newpage
Although we obtained our results using a specific set of Boolean update functions (eq. \ref{eq:update_function}), other Boolean threshold functions which explicitly include the current node value ($s_i(t)$), e.g. 
\begin{equation*}
s_i(t+1) =
    \begin{cases}
    1, & \sum_{j=1}^N G_{ij} s_j(t) > 0 \\
    s_i(t), & \sum_{j=1}^N G_{ij} s_j(t) \leq 0 \\
    \end{cases}
\end{equation*}
can be chosen and yield similar results (data not shown).

Most real-life datasets are incomplete (i.e., they do not show all possible attractors). The challenge is that the percentage of available attractors is unknown. Here we find a relationship between the percentage of known attractors and the average accuracy achieved in this evolutionary ESABO (Figure \ref{fig:Fitness_22nodes} and S2). This suggests the possibility to estimate the completeness of a set of microbiome abundance patterns. 
With more and more abundance patterns becoming available, this limitation will become less severe, but even now, as we have illustrated, our method can provide a rough estimate of how comprehensive the current data are.

Our findings prompt further research to understand in more detail the change of attractors under small variation of the underlying regulatory network, as a deeper understanding of this relationship has the potential of contributing better algorithms for the inference of microbial interaction networks. \bigskip \\

\section*{Acknowledgements}
This work was supported by the Deutsche Forschungsgemeinschaft (DFG, German Research Foundation) - GRK1657.

\newpage
\appendix

\section{Analytical formula for the calculation of $\mu$ and $\sigma$}
\label{sec:appendix_zScore_calc}

In order to speed up the computation of the ESABO scores $Z$ and to reduce the randomness of the obtained results,
%(occuring due to the random reshuffling of the original abundance vectors during the z-score calculation)
we introduce an analytical formula for the calculation of the mean $\mu$ and the standard deviation $\sigma$ of the entropy distribution, which is obtained if we calculate the entropy for all possible permutations $\pi(j)$ of the entries of $\vec{b}_j$.

The standard deviation can be calculated by
\begin{equation}
    \sigma= \langle H^2 \rangle-\mu^2
\end{equation}

with the mean
\begin{equation}
 \begin{split}
     \mu  &= \frac{1}{N_A!} \sum_{\pi \in P} H \left( \vec{x}_{i\pi(j)}^\text{AND}\right)\\
          &= \frac{1}{N_A!}  \sum_z H_z \, w(z)
 \end{split}
 \end{equation}
 
 and 
 \begin{equation}
     \langle H^2 \rangle = \frac{1}{N_A!}  \sum_z H_z^2 \, w(z). 
 \end{equation}

$N_A$ is the number of samples, respectively attractors,
$\pi(j)$ is a permutation of the entries of $\vec{b}_j$ %(''reshuffling``) 
and $P$ is the set of all possible permutations of the entries of $\vec{b}_j$. \\
$z(\pi(j))$ is the number of ones in $\vec{x}_{i\pi(j)}^\text{AND}$ and
$w(z_0)$ is the number of permutations $\pi$ that result in $z(\pi)=z_0$. \\

The number $w(z)$ of permutations that result in $z$ ones in $\vec{x}_{i\pi(j)}^\text{AND}$ can be calculated by 
 \begin{equation}
    w(z)=  \frac{n!}{(n-z)! } \frac{(N_A-n)!}{(N_A+z-n-m)!} {m\choose z} (N_A-m)!,
 \end{equation}
if  $z \in \left[\text{Max}(0,n+m-N_A),\text{Min}(n,m)\right]$. \\
Otherwise $w(z)=  0$. \\
 
$n=p_1(\vec{b}_i) \cdot N_A$ is the number of ones in $\vec{b}_i$ and $m=p_1(\vec{b}_j) \cdot N_A$ is the number of ones in $\vec{b}_j$ .

\section{Generation of ROC curves for the ESABO method and calculation of true positive and false positive rates for inferred or evolved networks}
\label{sec:appendix_ROC}

In order to create ROC curves for the ESABO method, we merge the link predictions for all the investigated networks into one large set and rank them according to their ESABO score.
For the recognition of positive links, the link predictions are sorted by descending ESABO score $Z$ and we only set positive links.
For the recognition of negative links, the link predictions are sorted by ascending ESABO score and we only set negative links.
For the recognition of links in general (regardless of their correct sign), the link predictions are sorted by descending absolute ESABO-score value $|Z|$. \\

The true positive rate (TPR) and false positive rate (FPR) of each evolved or inferred network 
is calculated by
\begin{equation}
    TPR_+ =  \frac{n_{++}}{L_{+}}, \quad
    FPR_+ =  \frac{n_{+-}+n_{+0}}{\frac{N\cdot (N-1)}{2}-L_{+}}
\end{equation}
    for the recognition of positive links, by \\
\begin{equation}
    TPR_- =  \frac{n_{--}}{L_{-}}, \quad
    FPR_- =  \frac{n_{-+}+n_{+0}}{\frac{N\cdot (N-1)}{2}-L_{-}}
\end{equation}
for recognition of negative links, and by
    \begin{align}
       &TPR_\text{edge} =  \frac{n_{++}+n_{--}+n_{+-}+n_{-+}}{L_{+}+L_{-}}, \nonumber \\ 
    &FPR_\text{edge} =  \frac{n_{+0}+n_{-0}}{\frac{N\cdot (N-1)}{2}-(L_{+}+L_{-})}
    \end{align}
for the recognition of edges in general. \\
$L_+$ is the number of positive edges and $L_-$ the number of negative edges in the original network. $N$ is the number of nodes. 
In the abbreviation $n_{xy}$, $y$ stands for the actual type of edge (positive (+), negative (-) or none (0)) and $x$ for the predicted relationship.
Hence, $n_{++}$ (respectively $n_{--}$) is the number of positive (resp. negative) edges that were correctly classified as positive (resp. negative). $n_{+-}$ (resp. $n_{-+}$) refers to the number of negative (resp. positive) edges that were wrongly classified as positive (resp. negative) and $n_{+0}$ (resp. $n_{-0}$) is the number of positive (resp. negative) edges that were present in the evolved or inferred network although there was no corresponding edge in the original network.

% \bibliographystyle{elsarticle-harv} 
 \bibliographystyle{elsarticle-num-names} 
 \bibliography{refs}





% \end{thebibliography}
\end{document}